\documentclass[12pt]{article}
\pdfoutput=1

\usepackage{amsmath, amsfonts, amssymb}
\usepackage{graphicx}
\usepackage{psfrag}
\usepackage[usenames,dvipsnames,svgnames,table]{xcolor}
\usepackage{enumerate}
\usepackage{jheppubcustom}
\usepackage{soul}
\usepackage{subfig}

\newcommand{\figref}[1]{Fig. \ref{#1}}

\renewcommand{\thefootnote}{\fnsymbol{footnote}}
\renewcommand{\thanks}[1]{\footnote{#1}} 
\newcommand{\starttext}{
\setcounter{footnote}{0}
\renewcommand{\thefootnote}{\arabic{footnote}}}

\newcommand{\be}{\begin{equation}}
\newcommand{\bea}{\begin{eqnarray}}
\newcommand{\eea}{\end{eqnarray}}
\newcommand{\beq}{\begin{equation}}
\newcommand{\ee}{\end{equation}}
\newcommand{\eeq}{\end{equation}}

\def\ba{\begin{eqnarray}}
\def\ea{\end{eqnarray}}

\def\12{{1 \over 2}}
\def\eq{&=&}

\def\simleq{\; \raise0.3ex\hbox{$<$\kern-0.75em
\raise-1.1ex\hbox{$\sim$}}\; }
\def\simgeq{\; \raise0.3ex\hbox{$>$\kern-0.75em
\raise-1.1ex\hbox{$\sim$}}\; }

\def\O2{\Omega_2}

\def\bi{\begin{itemize}}
\def\ei{\end{itemize}}

\def\sc{\setcounter{equation}{0}}

\def\W{$\Omega$}
\def\W'{$\Omega$}

\def\V{\Omega}
\def\V'{\Omega}

\def\lll{low-$l$}
\def\llls{low-$l$ suppression}

\title{Observational Consequences of a Landscape:  Epilogue}
\author[\spadesuit]{Ben Freivogel,} 
\author[\heartsuit]{Matthew Kleban,} 
\author[\clubsuit]{Mar\'ia Rodr\'iguez Mart\'inez,}
 \author[\diamondsuit]{and Leonard Susskind}

\emailAdd{benfreivogel@gmail.com, kleban@nyu.edu, mrm@zurich.ibm.com, susskind@stanford.edu}

\affiliation[\spadesuit]{\it GRAPPA and ITFA, Universiteit van Amsterdam,
Amsterdam, the Netherlands} 

\affiliation[\heartsuit]{\it Center for Cosmology and Particle Physics,
New York University, New York, USA}

\affiliation[\clubsuit]{\it IBM Research Z\"urich, 
S\"aumerstrasse 4, 8803 R\"uschlikon, Switzerland}

\affiliation[\diamondsuit]{\it Stanford Institute for Theoretical Physics and Department of Physics, 
Stanford University,
Stanford, USA}

\begin{document}

\bigskip\bigskip
\bigskip\bigskip
\abstract{

In this follow-up  to  \cite{Freivogel:2005vv} we briefly discuss the implications of the apparent detection of  $B$-modes in the Cosmic Microwave Background for the issues raised in that paper.
We argue that under the assumptions of eternal inflation, there is now stronger support for the detectability of a  Coleman-De Luccia bubble nucleation event in our past.  In particular, the odds that the spatial curvature of the universe is large enough to be detectable by near future experiments are increased.
}

\maketitle

\starttext \baselineskip=17.63pt \setcounter{footnote}{0}


\sc
\section{Review}

The purpose of this note is to discuss the relevance of the detection of CMB tensor modes \cite{Ade:2014xna} for a pattern that was  speculated on in \cite{Freivogel:2005vv} and in \cite{Bousso:2013uia}. The hypothetical pattern can be stated by three properties: convexity,  steepening of the inflaton potential, and suppression of scalar fluctuations at large angular scales (low $l$). These in themselves would be interesting properties of the inflaton potential, but the overall pattern, if confirmed, would  suggest something more far-reaching: namely, that the our part of the universe was born in a tunneling event from an earlier vacuum.  Large tensor power suggests that the tunneling event has a higher probability of detection than was previously estimated.

In calling attention to the pattern we feel that at the present time it is better to concentrate on  qualitative features without trying to be too quantitative about specific numerical models, or about theoretical  assumptions concerning  the probability measure for various parameters.

\bigskip

\noindent In \cite{Freivogel:2005vv} three related points were made. Let us review them.

\paragraph{Pressure toward shorter inflation.}
It is often stated that the reason for inflation is to flatten the universe. The cause-and-effect relationship is that inflation caused flatness; not that flatness caused inflation. This leaves us with the question of why inflation took place. The question is non-trivial because inflation is not a generic behavior; it typically requires a degree of fine-tuning. The answer offered in \cite{Freivogel:2005vv} was based on the assumption that our universe was born in a Coleman-De Luccia  (CDL) tunneling event from an earlier vacuum \cite{Coleman:1980aw}.

The product of a CDL tunneling event is an open universe, meaning that spatial slices are negatively curved. In a universe with negative spatial curvature, matter recedes with a velocity that on average is greater than escape velocity. Unless something is done to dilute the negative curvature, this outward velocity prevents structure formation, leading to an empty universe.  The ingredient that can dilute the curvature is a period of slow-roll inflation. Thus, the real rationale for inflation may be anthropic.

Anthropic considerations lead to a lower bound on the number of inflationary  e-foldings $N$  that took place after  the CDL event. At the same time there is also an observational lower bound on $N.$ The main result of  \cite{Freivogel:2005vv} is that the two lower bounds, anthropic and observational, only differ by about $2.5$ e-foldings. If we nominally say that the observational bound is $N\geq 60,$ then with the same conventions\footnote{The number of efoldings is significantly uncertain due to uncertainties about reheating after inflation. This uncertainty is irrelevant for the discussion here: it is always true that the anthropic bound is of order 2.5 efoldings away from the observational bound.} the anthropic bound is $N \geq 57.5.$

If anthropic constraints create pressure for a large value of $N,$ fine-tuning constraints create pressure for a small value of $N.$ The competition may lead to a situation where $N$ is close to the observational bound. This, in turn, could lead to observational signals from a tunneling event in our past.
We believe that the recent detection of tensor power at a surprisingly large magnitude bears on this question,  and increases the likelihood of such a signal.

\paragraph{Detectable spatial curvature?}
The signal we have in mind is the spatial curvature of the universe. A CDL tunneling event generally leads to negative curvature which may or may not be large enough to detect. The detection of negative curvature can be regarded as evidence that slow roll inflation began by bubble nucleation inside a ``parent vacuum"  different from our own. A detection of positive curvature would rule out any simple CDL mechanism.

The curvature parameter $\left| \Omega_k \right|$ is currently bounded to be less than about $.01$ in magnitude. On the other hand curvature at the level of $\simleq 10^{-5}$ would not be detectable due to cosmic variance.  It seems that there is a window of about two orders of magnitude in which observations of curvature would in principle  be possible. It is hard to overemphasize the importance that a detection of curvature---positive or negative---would have for cosmology. \cite{Kleban:2012ph, Guth:2012ww}

Another way to put it is in terms of $N.$ The window of opportunity for detecting curvature is between the observational bound and about three additional e-foldings. That is a fairly narrow window, and indeed, a simple statistical model of the parameters of inflation gave a probability of about $10 \%$ to be in that window (see \cite{othercurv} for a more detailed analysis of the probability.) The main point of this note is that the recent tensor-mode observations may significantly increase the likelihood to be in the window.

\paragraph{Effects on the power spectrum.} The third point made in \cite{Freivogel:2005vv} and more fully developed in \cite{Bousso:2013uia} is that if the number of e-foldings is near the observational bound, then a CDL origin may lead to a low $l$ suppression of the scalar fluctuation spectrum. The mechanism will be reviewed in section 3.

 The shape of the potential is correlated with the number of e-foldings that occurred between tunneling and the observable region. Consider the two potentials in \figref{1} and \figref{2}. They both represent landscapes that include tunneling events. The first figure is what we might call an optimistic case. The tunneling event is relatively close in field space to the value associated with the lowest values of $l.$  In this case a small number of e-foldings would separate the tunneling event from the observable region.

        By contrast the second figure is pessimistic; there are many e-foldings  separating tunneling and observation. While the overall shapes of the two cases are similar, in the observable region the shapes are quite different. In particular in the first case the observable low-$l$ physics took place on a convex portion of the curve while in the second case the potential was concave during the low-$l$ era.

        It was observed in \cite{Freivogel:2005vv} that the two cases lead to different observable effects on the low-$l$ scalar fluctuation spectrum. In particular, if the steepening of the potential in  \figref{1} is over a narrow range, it leads to a potentially large suppression of the low-$l$ scalar spectrum.

\subsection{Implications of new data}
\begin{figure}[h!]
\begin{center}
\includegraphics[scale=.5]{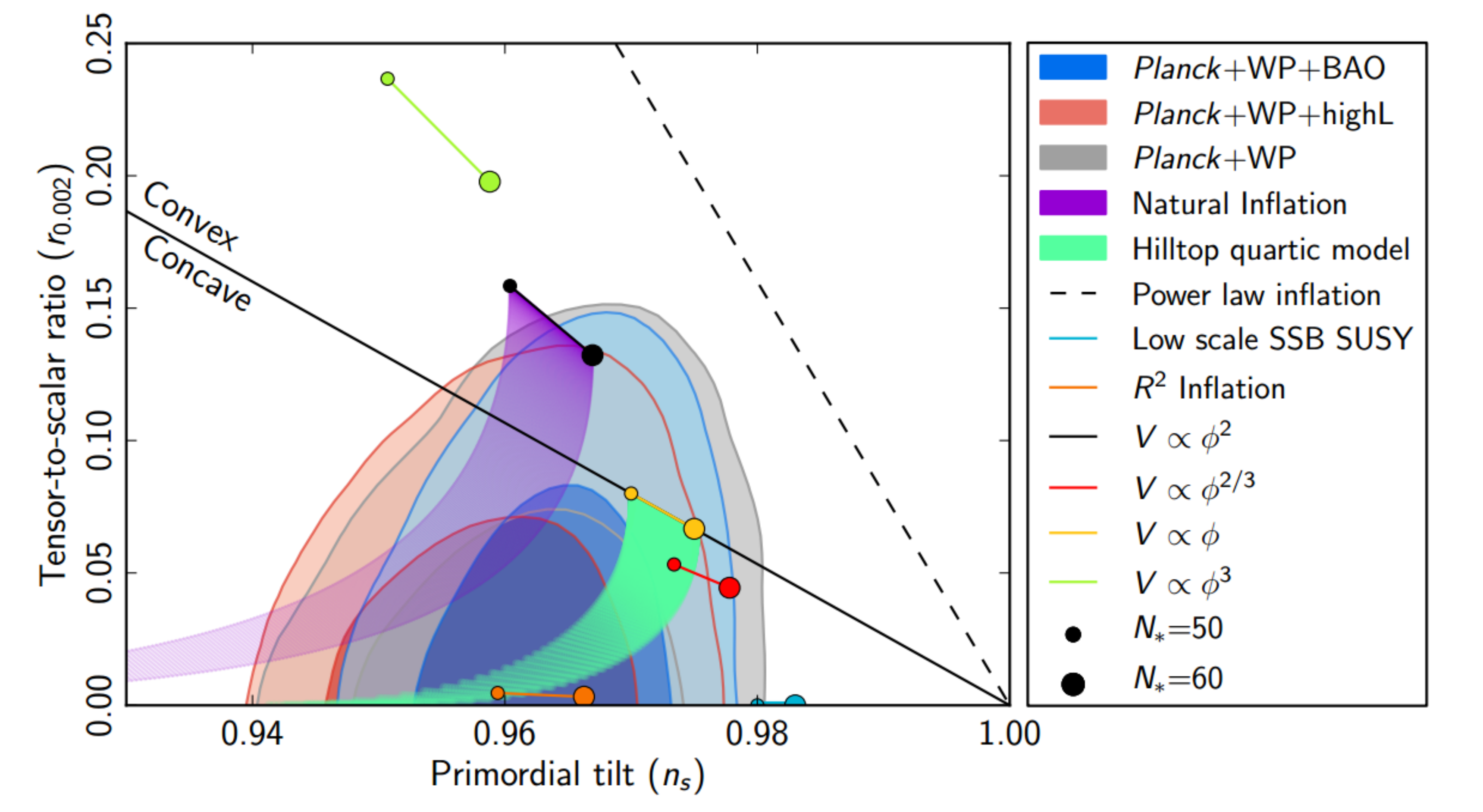}
\caption{This figure from \cite{Ade:2013uln} shows that a large tensor-scalar ratio of order  $.2$ favors
a convex potential in the observable region.}
\label{3}
\end{center}
\end{figure}

When \cite{Freivogel:2005vv} was written there was no  evidence for a systematic low-$l$ suppression, but that has changed significantly, first with WMAP and Planck \cite{Ade:2013nlj}, and apparently now even more-so with BICEP \cite{Ade:2014xna}; as emphasized in
\cite{Smith:2014kka} . The current situation is that a convex potential is favored and the power in temperature fluctuations at large angular scales is below the expectation based on the high-$l$ data, even more so if $r \approx .2$.  \figref{3} shows that a tensor-to-scalar ratio of $r\approx 0.2 $ lies well into the convex region.

\begin{figure}[h!]
\begin{center}
\includegraphics[scale=.65]{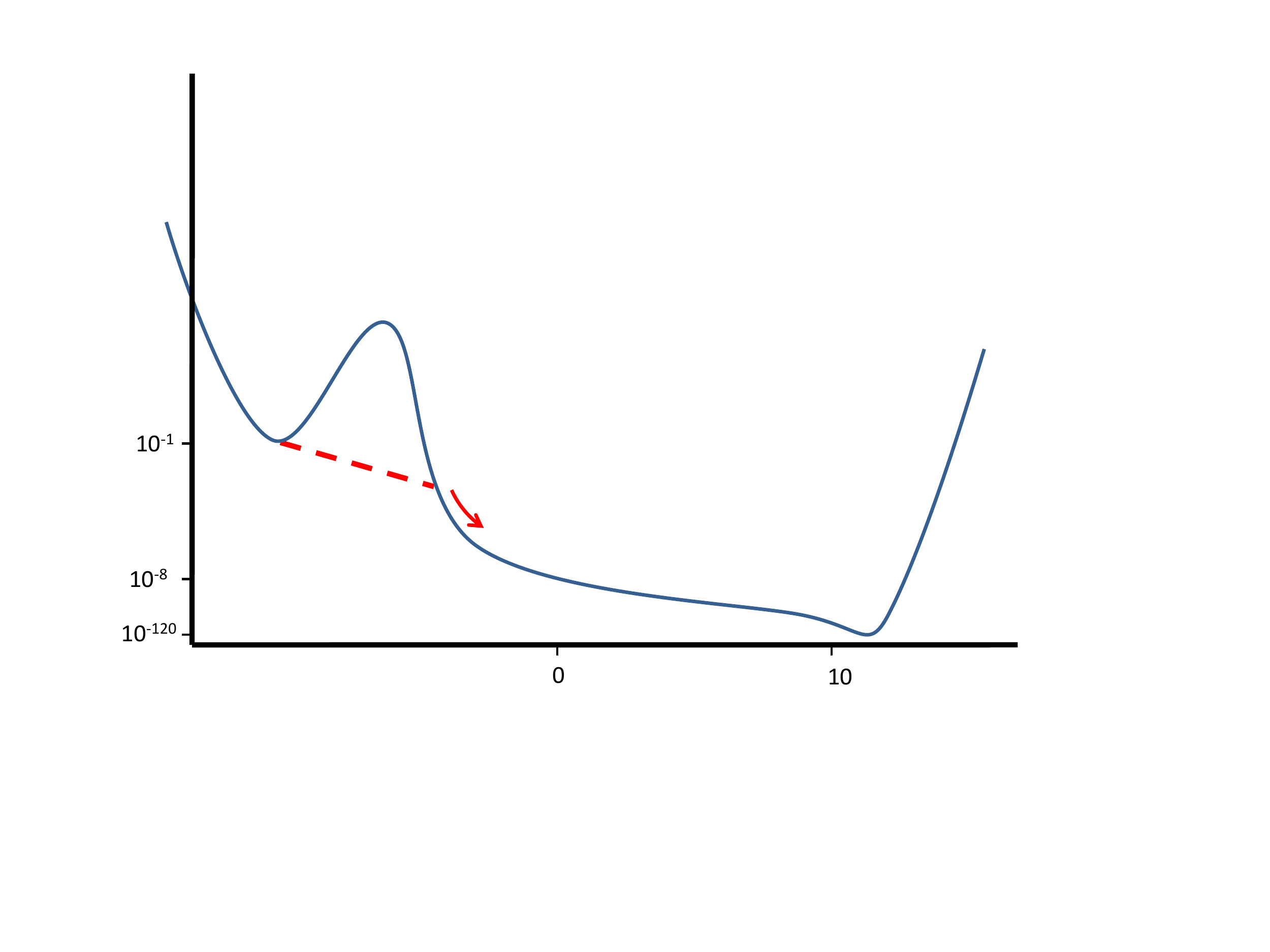}
\put(-430,320){\Large $ \frac{V}{m_{P}}$}
 \put(-105,90){\Large $ \frac{\phi}{m_{P}}$}
 \put(-280,145) {\large $H \sim V^{1/2} \sim 10^{-4}$}
\caption{Conjectured inflaton potential including a tunneling and slow-roll region, with $V$ and $\phi$ in units of the reduced Planck mass $m_P$. The figure is similar to one from
\cite{Freivogel:2005vv} and illustrates the optimistic case in which the tunneling event is close to the observational
region.  The point $\phi=0$ corresponds to a few efolds after the beginning of slow-roll inflation; the potential is steeper for $\phi<0$, therefore suppressing scalar power at large scales. 
}
\label{1}
\end{center}
\end{figure}

\begin{figure}[h!]
\begin{center}
\includegraphics[scale=.65]{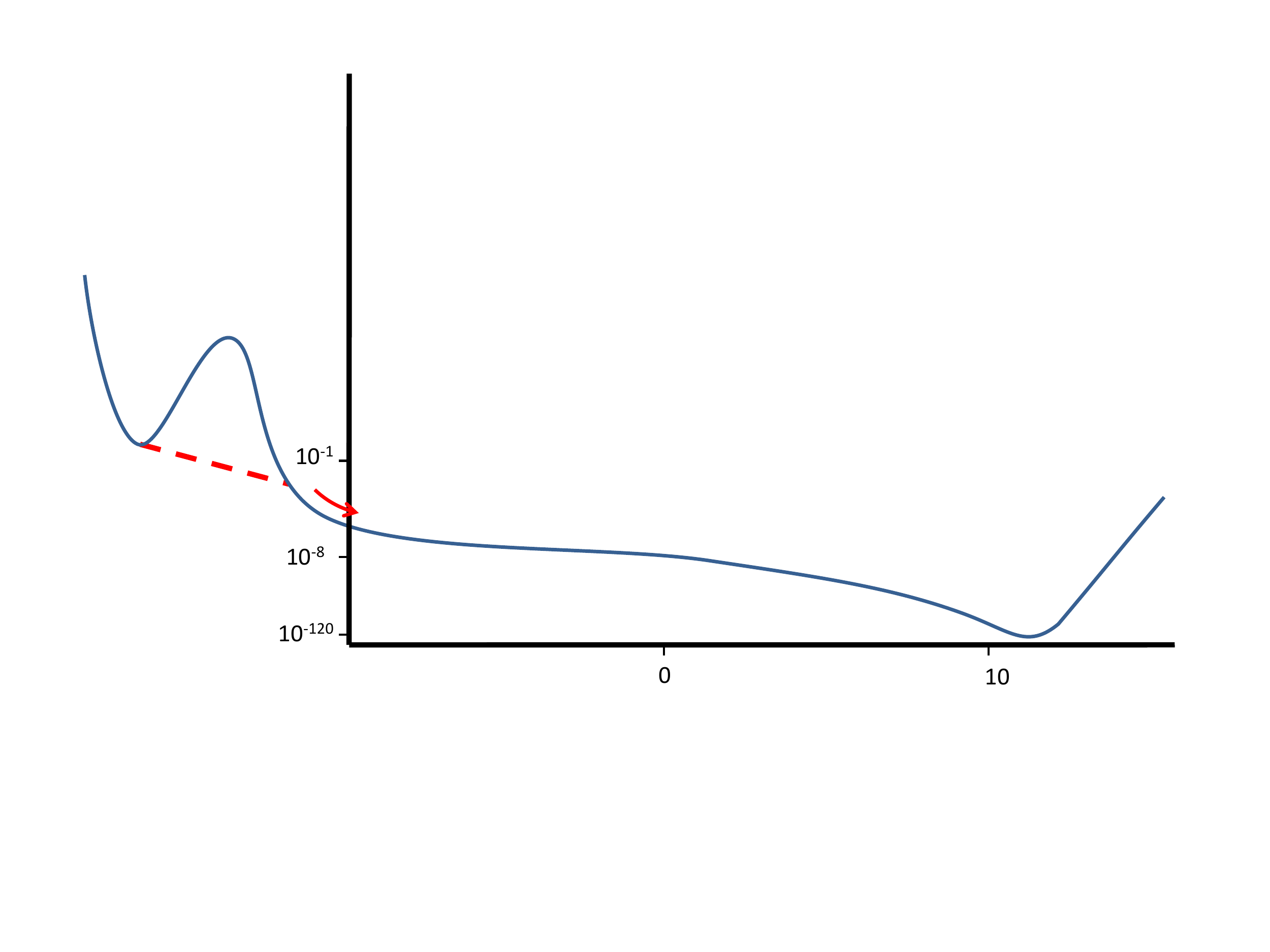}
\put(-370,320){\Large $ \frac{V}{m_{P}}$}
 \put(-60,95){\Large $ \frac{\phi}{m_{P}}$}
 \put(-265,157) {\large $H \sim V^{1/2} \sim 10^{-4}$}
\caption{The pessimistic case in which tunneling is separated from the observational
region ($\phi \sim 0$) by many e-foldings.  }
\label{2}
\end{center}
\end{figure}

A very interesting point was noticed in \cite{Bousso:2013uia}. Although the mechanism of \cite{Freivogel:2005vv} suppresses scalar power at low $l,$ it has no such effect on the tensor power. In fact it can even enhance the tensor power at low $l.$ Of course until power was seen in tensor modes, this was an academic point.

The observation of tensor modes at a level of  $r\approx 0.2 $   has an indirect implication for the shape of the potential. It means that some of the low-$l$ spectrum seen by earlier experiments is coming from tensors, and should be subtracted from the scalar power. In other words the low-$l$ suppression hinted at by Planck is stronger than was thought. Thus the low-$l$ data may have originated on a steeper slope. The evidence for a potential like  \figref{1}, as opposed to  \figref{2}, is strengthened, and that also implies that the number of e-foldings  between tunneling and observation may be small.

The main point of this epilogue is to point out that the pattern---convex potential; suppressed low-$l$ scalar power; small number of e-foldings---increases the likelihood that curvature may be detectable.

\paragraph{Relation to other work.} While this paper was in preparation we received a related work \cite{Hazra:2014aea} and became aware of \cite{bhs}. The observational effects of inflation that begins by bubble nucleation were thoroughly discussed by \cite{lindeetal}; see citations therein for important earlier work. An alternative explanation of the low $l$ anomalies is given in \cite{mattselfcite} and \cite{westped}. An interesting discussion of low $l$ anomalies in an open universe is given in \cite{liddle}.

\sc
\subsection{Why is Curvature So Important}

The importance of confirming or falsifying evidence for  a diverse landscape of vacua, and transitions between such vacua, can hardly be questioned. But there are obvious limitations on obtaining such evidence. In particular we are limited by the existence of a cosmological horizon which isolates us from most of the universe.
Nevertheless, we can  look back to the past, and hope to detect a transition from another vacuum. In other words we can hope to detect the fossil remnants of a CDL bubble nucleation, but only if the number of e-foldings is not too large. There is one unambiguous consequence of a CDL event: it leads to an open (negatively curved) cosmology.

A detection of curvature -- say at the level of $10^{-3} $ or $10^{-4}$ -- would be a game-changer \cite{Kleban:2012ph,Guth:2012ww}. Positive curvature would probably not completely end discussion about a multiverse but it would be very bad news for the eternal inflation/CDL bubble nucleation framework. On the other hand, a detection of negative curvature might not convince a skeptic, but it would be strong support for the CDL tunneling origin of our region  \cite{Kleban:2012ph,Guth:2012ww}.

In the rest of this epilogue we will review that argument for the connection between tunneling type potentials similar to  \figref{1} and the suppression of low-$l$ scalar power.

\sc
\section{Tunneling and Suppression}

In this section we will review the basic arguments connecting steepening and   \llls. It has been widely understood that a tensor-scalar ratio of order $.2$ requires a large inflaton excursion of $10$ or more Planck masses~\cite{Lyth:1996im}. In the notation of \cite{Freivogel:2005vv}:

\bi
\item The inflaton field is $\phi.$

\item The total excursion of $\phi$ from the onset of inflation until reheating is $\Delta \phi.$

\ei

In units with the reduced Planck mass $m_P^{2} \equiv \hbar c/8 \pi G = 1$, the observed  value of $r$ requires $\Delta \phi \simgeq 10 $.
On the other hand the value of the potential $V(\phi)$ over the \lll \ range is about $10^{-8}$ in the same units.  Just to have a concrete example, we consider a convex quadratic potential of the form,

\be
V(\phi) = \mu^2 (\phi - 10)^2
\label{quadpot}
\ee
with $\mu^2 = 10^{-10}.$ (Note that using literally this potential will lead to some issues, like a number of efoldings on the lower end of the allowed parameters and a rather red tilt. Our purpose here is simply to illustrate the effect of steepening, and (\ref{quadpot}) could be replaced with another potential.)

The lowest $l$ modes originate in the region close to $\phi \sim 0.$ In that range $|V'/V| \sim .2$  and the potential can be parameterized in the form given in \cite{Freivogel:2005vv}.

\bea
V \approx V_0 (1-.2 \phi)
\eea
where $V_{0}=100 \mu^{2} \sim 10^{-8}$.

However, this potential does not contain the characteristic feature of steepening that we expect if there were a nearby drop from a tunneling potential, as in  \figref{1}. In order to represent that behavior we can add in an additional ``steepening'' term
\bea \label{dV}
\delta V \eq \begin{cases}  - m \, \phi^3  \ \ \ \ \  &(\phi<0) \\
 0   \ \ \ \ \ \ \  &(\phi>0) \end{cases}
 \label{deltav}
\eea
that contributes for  $\phi<0$ but which vanishes for  $\phi > 0$ (we chose $\phi^{3}$ to make the potential continuous up to its second derivative, but other powers give very similar results).  This is an optimistic choice in that the steepening is assumed to occur in the same region as where the \lll \ modes leave the horizon.

More generally, the question of what potentials are natural in the context of the string landscape requires further study. The simplest explanation of the data is that the inflaton potential is very smooth over a surprisingly large field range, greater than the Planck mass. Such a smooth potential requires an approximate symmetry to protect it from generic corrections of the form $(\phi/M_P)^n$. Such potentials have been challenging to construct in string theory; some notable constructions are \cite{constructions}. Once we have such a symmetry allowing for a boring inflaton potential over a large range of $\phi$, it is not clear whether it is natural to have a steep potential from tunneling. One possibility is that the tunneling occurs from another direction in field space which is steeper. This is an interesting scenario that deserves further study, and it may lead to similar phenomenology as the simpler model considered here. 

For the moment, we restrict ourselves to the simpler single field model and compute the scalar fluctuation spectrum $\delta \rho$ as a function of $\phi,$
\be
\frac{\delta \rho}{\rho} =  \alpha \frac{V^{3/2}}{|V'|},
\label{basic}
\ee
where $\alpha = 1/(2 \pi \sqrt{3})$.
In the region of positive $\phi$ this is a nearly scale-invariant spectrum with a slight red tilt. But if we incorporate the steepening term, then for $\phi<0$ (\ref{basic}) gives

\be \label{sp}
\frac{\delta \rho}{\rho}= \alpha{\left[ \mu^2 (\phi-10)^2 + m^2 \phi^3\right]^{3/2} \over 2 \mu^2(10 - \phi) - 3 m^2 \phi^2} = \left({\delta \rho \over \rho}\right)_0 \left(1 -  {3 m^2 \phi^2 \over 20 \mu^2} + {\cal O}(\phi^{3})\right)
\ee
where the prefactor $\left({\delta \rho \over \rho}\right)_0 $ is the density perturbation without the steepening term. The negative term proportional to $m^2 \phi^2$ demonstrates the suppression in power due to the steepened potential. Recall that in our conventions the minimum of the potential is at $\phi = 10$.


\begin{figure}[h!]
\begin{center}
\includegraphics[scale=.65]{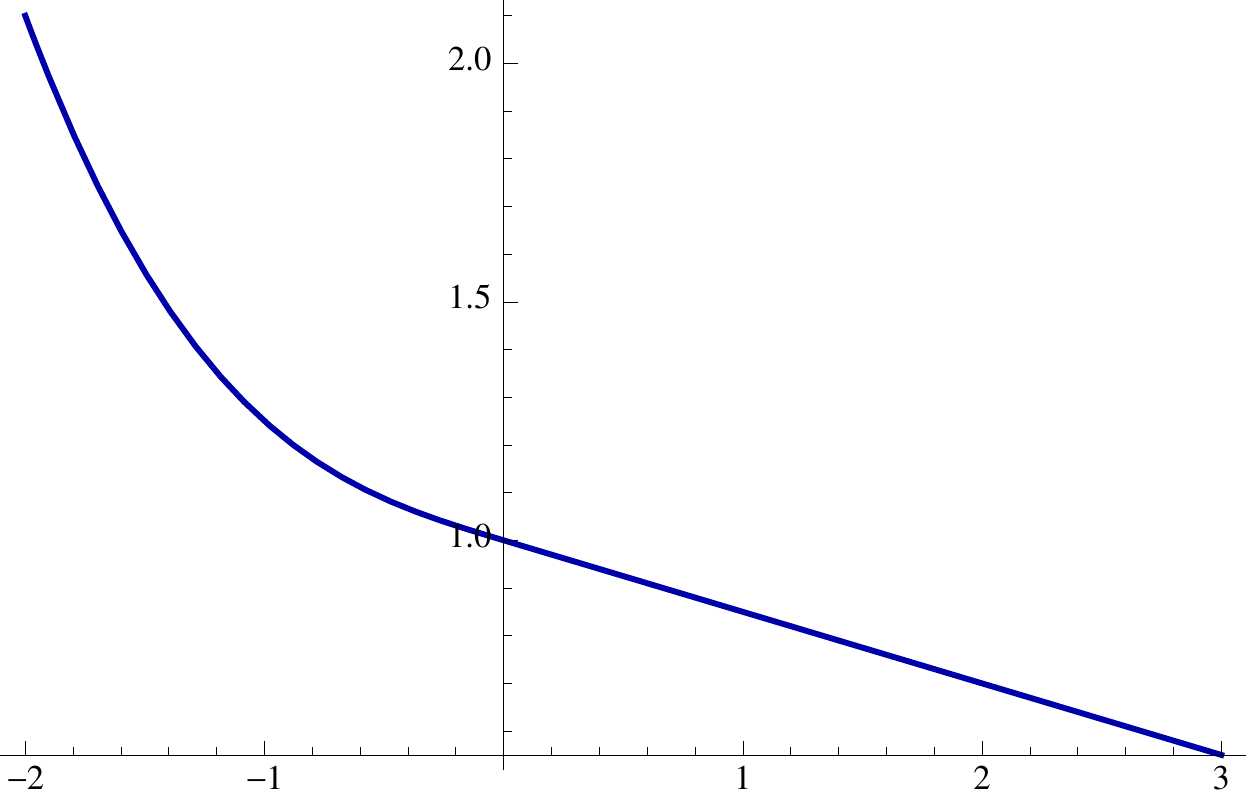}
\put(-170,160){\large ${V \times 10^{8}/ m_{P}^4}$}
 \put(5,5){\Large $ \frac{\phi}{m_{P}}$}
\caption{An example of an inflation potential $V(\phi)$, defined in equations (\ref{quadpot}) and (\ref{deltav}) in the vicinity of $\phi=0$. This potential is of the type illustrated in figure 1.  }
\label{4}
\end{center}
\end{figure}

\begin{figure}[h!]
\begin{center}
\includegraphics[scale=.65]{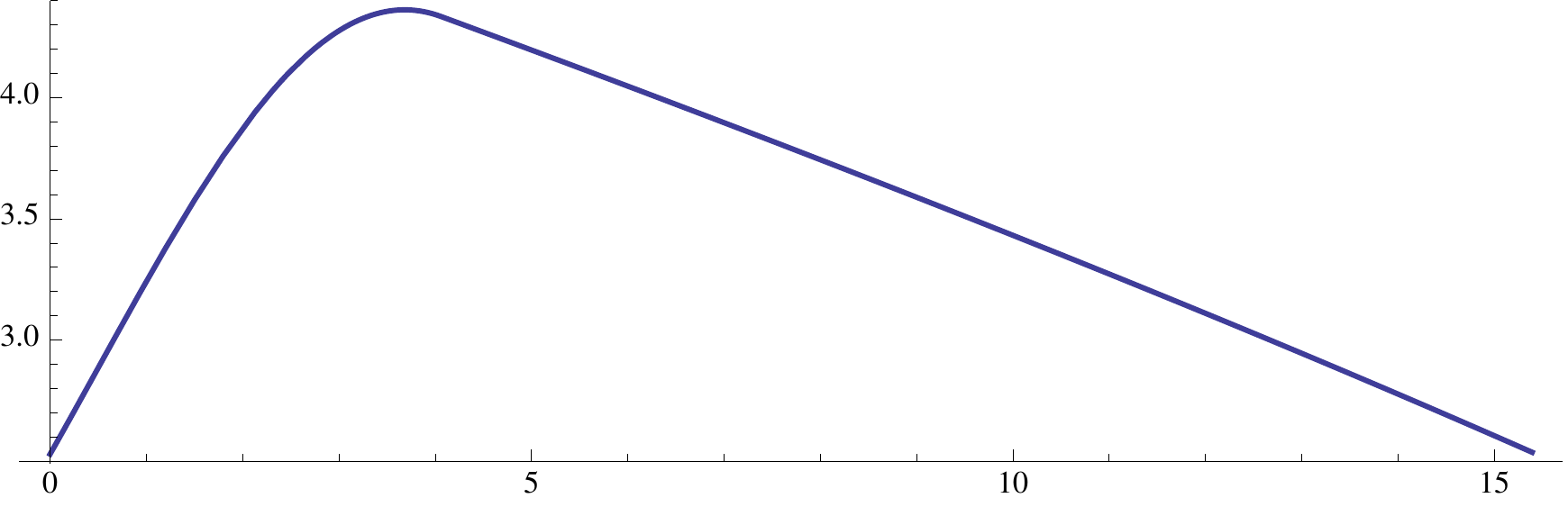}
\put(-345,110){\large ${\delta \rho/\rho} \times 10^5$}
 \put(5,5){\large $N-N_Q$}
\caption{Scalar fluctuations calculated in the slow-roll approximation as a function of the number of efolds $N$ from the quadrupole $N_Q$, for the potential plotted in \figref{4}.  For this example scalar power is suppressed for $l \simleq 50$, while primordial tensor power is close to scale-invariant.  The point $N-N_Q=0$ on this plot has been chosen to correspond to $\phi/m_P \approx -.75$ in \figref{4}; $N-N_Q=15$ corresponds to $\phi/m_P \approx 2$.}
\label{5}
\end{center}
\end{figure}

To make an estimate of the range of $l$ over which steepening suppresses the scalar power, we assume that the CMB quadrupole modes leave the horizon when the inflaton has a value $\phi_0<0.$ Consider the number of e-foldings that take place during the time that the field rolls from $\phi=\phi_0$ to $\phi=0.$ Call that number $N_s(\phi_{0})$, where $s$ stands for suppressed.  If we choose $\phi_{0}$ so that $N_{s}\sim 4$, the power in scalar fluctuations will be  suppressed for $l \simleq e^{4} \approx 50$.  We can choose $m$ so that the tilt for that range of $l$ is blue, for instance $n_{s}-1 \approx 1.05$, which appears to be a good fit to the power suppression at low $l$\footnote{Raphael Flauger, private communication.}.  However this  example is just an illustration of the suppression mechanism between $\phi_0$ and $\phi =0,$ that is, between the lowest $l$ and the end of the suppression at $l \approx 50.  $ 

The choice $\delta V \sim \phi^{3}$ in (\ref{dV}) actually leads to an increase in the scalar power  for  $  \phi  \ll \phi_{0}$.      This increase at (unobservably) long wavelengths  is an artifact of this specific choice for the steepening.  With further steepening of the potential close to the tunneling point the increase of power can easily be avoided.

Depending on how many efolds of slow-roll inflation took place after the tunneling but before the field reached $\phi_{0}$, the curvature will be diluted to a greater or lesser extent.  If $\phi=\phi_{0}$ is the beginning of the post-tunneling slow-roll phase the universe would be curvature dominated today.  If there are $\sim 3$ efolds of inflation prior to $\phi=\phi_{0}$, then curvature would be $|\Omega_{k}| \approx e^{-2*3} \approx 2 \times 10^{-3}$.  More than $\sim 6$ efolds of inflation prior to $\phi=\phi_{0}$ would make $|\Omega_{k}| < 10^{-5}$ and therefore impossible to detect \cite{Kleban:2012ph}.


The point is not so much that a tunneling event in the past predicts a \llls, but rather that the observed \llls  \ suggests that the tunneling transition is close (in field space) to the value associated with the \lll\  power.

We could have parameterized the steepening term so that it turned on at some large negative $\phi$ as in  \figref{2}.  That would have  two effects. First it would have eliminated the \llls. Secondly it would imply a relatively large number of e-foldings between tunneling and the \lll \ region. As explained in \cite{Freivogel:2005vv} and \cite{Bousso:2013uia} the two go together. The optimistic view is that the observed \llls \ suggests that few e-foldings separate tunneling from observation.

Without trying to be quantitative about probability measures, the BICEP data in our view tend to increase the likelihood that curvature can be detected. Obviously every effort should be made to do so.

\section*{Acknowledgements}

It is a pleasure to thank Raphael Flauger, Matthew Johnson, David Spergel, and Matias Zaldarriaga for useful conversations.
Support for the research of LS came through NSF grant Phy-1316699 and the Stanford Institute for Theoretical Physics.  The work of MK is supported in part by the NSF  through grant PHY-1214302 and by the John Templeton Foundation.  The opinions expressed
in this publication are those of the authors and do not necessarily reflect the views of the John Templeton Foundation.


\begin{thebibliography}{99}

\bibitem{Ade:2014xna}
  P.~A.~R.~Ade {\it et al.}  [BICEP2 Collaboration],
  ``BICEP2 I: Detection Of B-mode Polarization at Degree Angular Scales,''
  arXiv:1403.3985 [astro-ph.CO].

\bibitem{Freivogel:2005vv}
  B.~Freivogel, M.~Kleban, M.~Rodriguez Martinez and L.~Susskind,
 ``Observational consequences of a landscape,''
  JHEP {\bf 0603}, 039 (2006)
  [hep-th/0505232].
  
 
\bibitem{Bousso:2013uia}
  R.~Bousso, D.~Harlow and L.~Senatore,
  ``Inflation after False Vacuum Decay: Observational Prospects after Planck,''
  arXiv:1309.4060 [hep-th].

 \bibitem{othercurv}
  J.~March-Russell and F.~Riva,
  ``Signals of Inflation in a Friendly String Landscape,''
  JHEP {\bf 0607}, 033 (2006)
  [astro-ph/0604254].
  
  A.~De Simone and M.~P.~Salem,
  ``The distribution of $\Omega_k$ from the scale-factor cutoff measure,''
  Phys.\ Rev.\ D {\bf 81}, 083527 (2010)
  [arXiv:0912.3783 [hep-th]].


\bibitem{Coleman:1980aw} 
  S.~R.~Coleman and F.~De Luccia,
  ``Gravitational Effects on and of Vacuum Decay,''
  Phys.\ Rev.\ D {\bf 21}, 3305 (1980).


\bibitem{Hazra:2014aea} 
  D.~K.~Hazra, A.~Shafieloo, G.~F.~Smoot and A.~A.~Starobinsky,
  ``Ruling out the power-law form of the scalar primordial spectrum,''
  arXiv:1403.7786 [astro-ph.CO].

\bibitem{bhs}R.~Bousso, D.~Harlow, and L.~Senatore, to appear.

\bibitem{lindeetal} 
  D.~Yamauchi, A.~Linde, A.~Naruko, M.~Sasaki and T.~Tanaka,
  ``Open inflation in the landscape,''
  Phys.\ Rev.\ D {\bf 84}, 043513 (2011)
  [arXiv:1105.2674 [hep-th]].
  
\bibitem{mattselfcite} 
  G.~D'Amico, R.~Gobbetti, M.~Kleban and M.~Schillo,
  ``Large-scale anomalies from primordial dissipation,''
  JCAP {\bf 1311}, 013 (2013)
  [arXiv:1306.6872 [astro-ph.CO]].
 
\bibitem{liddle} 
  A.~R.~Liddle and M.~Cortês,
  ``Cosmic microwave background anomalies in an open universe,''
  Phys.\ Rev.\ Lett.\  {\bf 111}, 111302 (2013)
  [arXiv:1306.5698 [astro-ph.CO]].
  
\bibitem{westped} 
  F.~G.~Pedro and A.~Westphal,
  ``Low-l CMB Power Loss in String Inflation,''
  arXiv:1309.3413 [hep-th].
  
 
  

\bibitem{Kleban:2012ph}
  M.~Kleban and M.~Schillo,
  ``Spatial Curvature Falsifies Eternal Inflation,''
  JCAP {\bf 1206}, 029 (2012)
  [arXiv:1202.5037 [astro-ph.CO]].

\bibitem{Guth:2012ww}
  A.~H.~Guth and Y.~Nomura,
  ``What can the observation of nonzero curvature tell us?,''
  Phys.\ Rev.\ D {\bf 86}, 023534 (2012)
  [arXiv:1203.6876 [hep-th]].

\bibitem{Ade:2013uln} 
  P.~A.~R.~Ade {\it et al.}  [Planck Collaboration],
  ``Planck 2013 results. XXII. Constraints on inflation,''
  arXiv:1303.5082 [astro-ph.CO].

\bibitem{Ade:2013nlj} 
  P.~A.~R.~Ade {\it et al.}  [Planck Collaboration],
  ``Planck 2013 results. XXIII. Isotropy and statistics of the CMB,''
  arXiv:1303.5083 [astro-ph.CO].
  
  \bibitem{Smith:2014kka} 
  K.~M.~Smith, C.~Dvorkin, L.~Boyle, N.~Turok, M.~Halpern, G.~Hinshaw and B.~Gold,
  ``On quantifying and resolving the BICEP2/Planck tension over gravitational waves,''
  arXiv:1404.0373 [astro-ph.CO].

  
\bibitem{Lyth:1996im} 
  D.~H.~Lyth,
  ``What would we learn by detecting a gravitational wave signal in the cosmic microwave background anisotropy?,''
  Phys.\ Rev.\ Lett.\  {\bf 78}, 1861 (1997)
  [hep-ph/9606387].
  
  \bibitem{constructions}
  N.~Kaloper, A.~Lawrence and L.~Sorbo,
  ``An Ignoble Approach to Large Field Inflation,''
  JCAP {\bf 1103}, 023 (2011)
  [arXiv:1101.0026 [hep-th]].
  
  N.~Kaloper and L.~Sorbo,
  ``A Natural Framework for Chaotic Inflation,''
  Phys.\ Rev.\ Lett.\  {\bf 102}, 121301 (2009)
  [arXiv:0811.1989 [hep-th]].

  X.~Dong, B.~Horn, E.~Silverstein and A.~Westphal,
  ``Simple exercises to flatten your potential,''
  Phys.\ Rev.\ D {\bf 84}, 026011 (2011)
  [arXiv:1011.4521 [hep-th]].
  
  L.~McAllister, E.~Silverstein and A.~Westphal,
  ``Gravity Waves and Linear Inflation from Axion Monodromy,''
  Phys.\ Rev.\ D {\bf 82}, 046003 (2010)
  [arXiv:0808.0706 [hep-th]].
  E.~Silverstein and A.~Westphal,
  ``Monodromy in the CMB: Gravity Waves and String Inflation,''
  Phys.\ Rev.\ D {\bf 78}, 106003 (2008)
  [arXiv:0803.3085 [hep-th]].

\end{thebibliography}
\end{document}